\documentclass[showpacs,preprintnumbers, amsmath,amssymb,superscriptaddress,preprint,floatfix]{revtex4}
\usepackage{graphicx}
\draft
\begin{document}
\newcommand {\bb}{\bibitem}
\newcommand {\be}{\begin{equation}}
\newcommand {\ee}{\end{equation}}
\newcommand {\bea}{\begin{eqnarray}}
\newcommand {\eea}{\end{eqnarray}}
\newcommand {\nn}{\nonumber}

\title{Gap Symmetry of Superconductivity in UPd$_{2}$Al$_{3}$}

\author{H. Won}

\address{Department of Physics, Hallym University,
Chuncheon 200-702, South Korea}

\author{D. Parker}
\author{K. Maki}

\address{Department of Physics and Astronomy, University of Southern
California, Los Angeles, CA 90089-0484 USA}

\author{T. Watanabe}
\author{K. Izawa}
\author{Y. Matsuda}

\address{Institute for Solid State Physics, University of Tokyo, Kashiwanoha 5-1-5,
Kashiwa,Chiba 277-8581, Japan}

\date{\today}

\begin{abstract}

The angle dependent thermal conductivity of the heavy-fermion superconductor
UPd$_2$Al$_3$ in the vortex state was recently measured by Watanabe et al.  Here we analyze this
data from two perspectives: universal heat conduction and
the angle-dependence.  We conclude that the superconducting
gap function $\Delta({\bf k})$ in UPd$_2$Al$_3$ has horizontal nodes and is given
by $\Delta({\bf k}) =\Delta\cos(2\chi)$, with $\chi = ck_{z}$.

\end{abstract}
\maketitle
 
\noindent{\it \bf 1.  Introduction}

Since the discovery of the heavy-fermion superconductor CeCu$_2$Si$_{2}$
in 1979\cite{1} the gap symmetries of unconventional superconductors have become 
a central issue in condensed-matter physics\cite{2}. 
In the last few years, the angle-dependent magnetothermal conductivity in the vortex state
of nodal superconductors has been established as a powerful technique to address the gap 
symmetry.  This is in part due to the theoretical understanding of the quasi-particle
spectrum in the vortex state of nodal superconductors, following the path-breaking work by
Volovik\cite{3,4,5}.  Using this approach, Izawa et al have succeeded in identifying the gap 
symmetries of superconductivity in Sr$_{2}$RuO$_{4}$, CeCoIn$_{5}$, $\kappa$-(ET)$_{2}$Cu(NCS)$_{2}$, 
YNi$_{2}$B$_{2}$C, and PrOs$_4$Sb$_{12}$\cite{6,7,8,9,10}. 

Superconductivity in UPd$_2$Al$_3$ was discovered by Geibel et \cite{11} in 1991.  The
reduction of the Knight shift in NMR\cite{12} and the Pauli limiting of H$_{c2}$\cite{13} indicate 
spin singlet pairing in this compound.  Nodal superconductivity with horizontal nodes has been
suggested from the thermal conductivity data \cite{14} and from the c-axis tunneling data of
thin film UPd$_2$Al$_3$ samples\cite{15}.  Very recently, McHale et al \cite{16} have proposed 
$\Delta({\bf k}) =\Delta\cos(\chi)$ (with $\chi = ck_{z}$) based on a model where the
pairing interaction arises from antiparamagnon exchange with ${\bf Q}=(0,0,\frac{\pi}{c})$
\cite{17}.  Furthermore, the thermal conductivity data of UPd$_{2}$Al$_{3}$ for a variety of
magnetic field orientations have been reported\cite{18}.  At first glimpse the experimental
data appeared to support the model proposed by McHale et al.

The object of the present paper is to show that an alternative model, i.e. 
$\Delta({\bf k}) =\Delta\cos(2\chi)$,
descibes the thermal conductivity data more consistently.  For this purpose we first
generalize the universal heat conduction initially proposed in the context of d-wave 
superconductivity\cite{19,20} to a variety
of nodal superconductors.  We limit ourselves to quasi-2D systems with $\Delta({\bf k}) =
\Delta f$ and $f=\cos(2\phi),\sin(2\phi),\cos\chi, e^{i \phi}\cos\chi, \cos(2\chi)$ 
$\sin\chi$, and $e^{i \phi}\sin\chi.$
It is found that the in-plane thermal conductivity $\kappa_{xx}$ is independent of f.  
On the other hand, the out-of-plane thermal conductivity $\kappa_{zz}$ can discriminate
different f's.  
Second, we extend an early study of the angle-dependent thermal conductivity\cite{21} for
$\kappa_{yy}$ in a magnetic field rotated in the z-x plane.  The comparison of these results with
experimental data indicates $\Delta({\bf k}) =\Delta\cos(2\chi)$.

\noindent{\it \bf 2. Universal Heat Conduction}

Here we consider the thermal conductivity $\kappa$ in the limit $T \rightarrow 0 K$ in the presence
of disorder.  It is assumed that the impurities are in the unitary scattering limit\cite{20}.
We consider the quasi-2D gap functions $\Delta({\bf k}) =\Delta f$ with
$f=\cos(2\phi),\sin(2\phi)$[d-wave superconductor as in the high-T$_{c}$ cuprates]
,$\cos \chi, e^{i \phi} \cos\chi$ (f-wave superconductor as proposed for Sr$_{2}$RuO$_{4}$ \cite{6}),
$\cos(2\chi)$, $\sin \chi$, and $e^{i \phi} \sin\chi .$  
Following Ref.\cite{20}, the thermal conductivity within the 
conducting plane is given by 
\bea
\kappa_{xx}/\kappa_{n}= \kappa_{yy}/\kappa_{n} &=& \frac{\Gamma_{0}}{\Delta} \langle 
(1+\cos(2\phi))\frac{C_{0}^{2}}{(C_{0}^{2}+|f|^{2})^{\frac{3}{2}}} \rangle \\
&=& \frac{2\Gamma_{0}}{\pi \Delta \sqrt{1+C_{0}^{2}}}E(\frac{1}{\sqrt{1+C_{0}^{2}}})
=I_{1}(\Gamma/\Gamma_{0})
\eea
where $\kappa_{n}$ is the thermal conductivity in the normal state when $\Gamma=\Gamma_{0}$, and $\Gamma$
is the quasi-particle scattering rate in the normal state.  Here $\langle .... \rangle$ denotes
the average over $\phi$ and $\chi$, and Eq.(1) tells us that the planar thermal conductivity
is independent of the gap functions given above.  
Also $\Gamma_{0}=\frac{\pi}{2\gamma}T_{c}= 0.866 T_{c}$ and $T_{c}$ is the superconducting
transition temperature of the pure system.  However, the quasi-particle scattering rate at E=0 is given
by $\Delta C_{0}$, and $C_{0}$ is determined by \cite{20}
\bea
\frac{C_{0}^{2}}{\sqrt{1+C_{0}^{2}}} K(\frac{1}{\sqrt{1+C_{0}^{2}}}) = \frac{\pi \Gamma}{2\Delta}
\eea
and $\Delta=\Delta(0,\Gamma)$ has to be determined self-consistently as in \cite{20}.  Here K(k) and
E(k) are the complete elliptic integrals.  We show $I_{1}(\Gamma/\Gamma_{0})$ in Fig.1.
\begin{figure}[h]
\includegraphics[width=6.75cm]{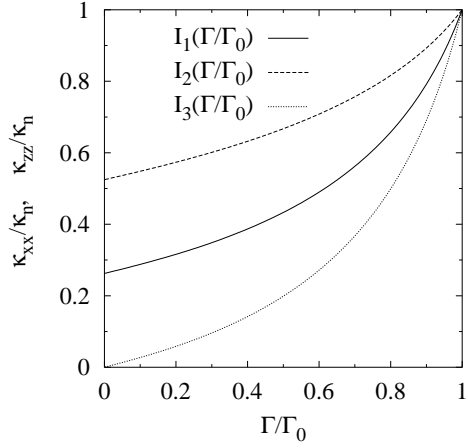}
\caption{The functions I$_{1}$, I$_{2}$ and I$_{3}$.}
\end{figure}
Now let us look at the out-of-plane thermal conductivity $\kappa_{zz}$.  This is given by
\bea
\kappa_{zz}/\kappa_{n} &=& \frac{\Gamma_{0}}{\Delta} \langle (1-\cos(2\chi))\frac{C_{0}^{2}}{(C_{0}^{2}+|f|^{2})^{\frac{3}{2}}} \rangle \\
&=& I_{1}(\Gamma/\Gamma_{0})
\eea
for $f=\cos(2\phi),\sin(2\phi)$ and $\cos(2\chi)$, but
\bea
\frac{\kappa_{zz}}{\kappa_{n}}&=& \frac{4\Gamma_{0}}{\pi \Delta \sqrt{1+C_{0}^{2}}}
\left(E(\frac{1}{\sqrt{1+C_{0}^{2}}})-C_{0}^{2} (K(\frac{1}{\sqrt{1+C_{0}^{2}}})-E(\frac{1}
{\sqrt{1+C_{0}^{2}}}))\right)\\
&=& I_{2}(\frac{\Gamma}{\Gamma_{0}}) 
\eea
for $f=\cos\chi, e^{\pm i \phi} \cos\chi$, and
\bea
\kappa_{zz}/\kappa_{n} = \frac{2\Gamma_{0} \Gamma}{\Delta} 
\left( 1 - \frac{E(\frac{1}{\sqrt{1+C_{0}^{2}}})}{K(\frac{1}{\sqrt{1+C_{0}^{2}}})}\right) \equiv 
I_{3}(\frac{\Gamma}{\Gamma_{0}})
\eea
for $f=\sin\chi, e^{i \phi} \sin\chi$.  These functions are shown in Fig. 1.

In Fig. 2 we show $\kappa_{yy}(H)$
and $\kappa_{zz}$ for ${\bf H} \parallel \hat{z}$ taken for UPd$_2$Al$_3$ \cite{22}.  In particular
\begin{figure}[h]
\includegraphics[width=6.75cm]{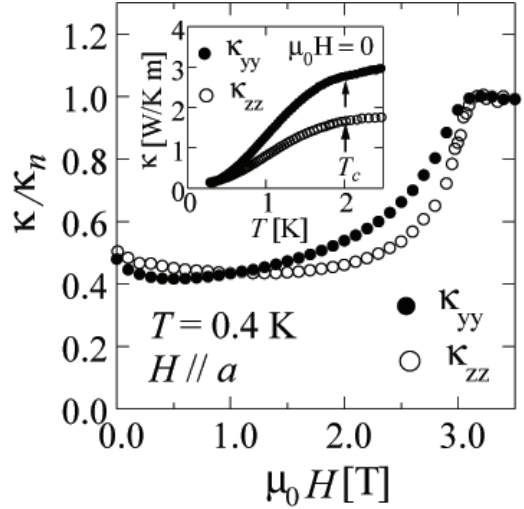}
\caption{$\kappa_{yy}(H)$ and $\kappa_{zz}(H)$ for UPd$_{2}$Al$_{3}$}
\end{figure}
$(\kappa_{00})_{yy}=(\kappa_{00})_{zz}$ indicates $\Delta({\bf k}) \sim \cos(2\chi)$.  Of course
the effect of the magnetic field is not equivalent to the effect of impurities.  But this
comparison points to $\Delta({\bf k}) \sim \cos(2\chi)$ for UPd$_2$Al$_3$.  We note also that
for $f=\sin\chi$ and $e^{i \phi} \sin\chi$, there will be no universal heat conduction in $\kappa_{zz}$.

\noindent{\it \bf 3. Angle-dependent magnetothermal conductivity}

First let us recapture the quasiparticle density of states in the vortex state of nodal
superconductors. \cite{4}  For simplicity we consider f's with horizontal nodes:
$f= e^{i \phi}\cos\chi, \cos \chi, \cos 2\chi$, $\sin \chi$ and $e^{i \phi} \sin\chi$\cite{21}.  
Then the first two
f's have nodes at $\chi_{0}= \pm \frac{\pi}{2}$, whereas $f=\cos 2\chi$ at 
$\chi_{0}= \pm \frac{\pi}{4}$ and $f = \sin \chi$ and $e^{i \phi} \sin\chi$at $\chi_{0}=0$.

In an arbitrary field orientation we obtain the quasiparticle density of states
\bea
\mathcal{G}({\bf H}) \equiv \frac{N(0,{\bf H})}{N_{0}} = \frac{2}{\pi^{2}}\frac{v_{a} 
\sqrt{eH}}{\Delta} I_{1}(\theta)
\eea
for the superclean limit and 
\bea
\mathcal{G}({\bf H}) \simeq (\frac{2\Gamma}{\pi \Delta})^{1/2} 
[\log(4\sqrt{\frac{2\Delta}{\pi \Gamma}})]^{1/2} 
(1 + \frac{v_{a}^{2}eH}{8 \pi^{2} \Gamma \Delta} \log(\frac{\Delta}{v_{a}\sqrt{eH}}) I_{2}(\theta))
\eea
for the clean limit, where
\bea
\nn
I_{1}(\theta)&=&(\cos^{2}\theta+\alpha \sin^{2}\theta)^{1/4} 
\frac{1}{\pi}\int_{0}^{\pi}d\phi\left(\cos^{2}\theta + \sin^{2}\theta (\sin^{2}\phi+\alpha \sin^{2}\chi_{0})+\sqrt{\alpha}\sin(\chi_{0})\cos \phi \sin(2\theta) \right) ^{1/2}\\
\nn
& \simeq & (\cos^{2}\theta + \alpha \sin^{2}\theta)^{1/4}(1+\sin^{2}\theta(-\frac{1}{2} + \alpha
\sin^{2}\chi_{0}))^{1/2} \left(1 - \frac{1}{64}\frac{\sin^{2}\theta(\sin^{2}\theta + 16 \alpha \sin^{2}\chi_{0}
\cos^{2}\theta)}{(1+\sin^{2}\theta(-\frac{1}{2}+\alpha \sin^{2}\chi_{0}))^{2}}\right)
\eea
and
\bea
I_{2}(\theta) &=& (\cos^{2}\theta + \alpha \sin^{2}\theta)^{1/2}(1+\sin^{2}\theta(-\frac{1}{2}+\alpha 
\sin^{2}\chi_{0})).
\eea
Here $\alpha=(v_{c}/v_{a})^{2}$ and $\theta$ is the angle ${\bf H}$ makes from the z-axis.  Then
the specific heat, the spin susceptibility and the planar superfluid density in the vortex state
in the limit $T \rightarrow 0 K$ are given by \cite{23}
\bea
C_{s}/\gamma_{N} T  &=& \mathcal{G}({\bf H}), \frac{\chi_{S}}{\chi_{N}}= \mathcal{G}({\bf H}),\\ 
\frac{\rho_{S\parallel}({\bf H})}{\rho_{S\parallel}(0)} &=& 1 - \mathcal{G}({\bf H})
\eea
Similarly the thermal conductivity $\kappa_{yy}$ when the magnetic field is rotated in the
z-x plane is given by 
\bea
\frac{\kappa_{yy}}{\kappa_{n}} &=& \frac{2}{\pi^{3}}\frac{v_{a}^{2}eH}{\Delta^{2}}F_{1}(\theta)
\eea
in the superclean limit and
\bea
\frac{\kappa_{yy}}{\kappa_{00}}&=& 1+ \frac{v_{a}^{2}(eH)}{6\pi^{2}\Gamma \Delta}F_{2}(\theta)
\log(2\sqrt{\frac{2\Delta}{\pi \Gamma}})\log(\frac{2\Delta}{v_{a}\sqrt{eH}})
\eea
in the clean limit where 
\bea
F_{1}(\theta) &=& \sqrt{\cos^{2}\theta+\alpha \sin^{2}\theta} (1 + \sin^{2}\theta(-\frac{3}{8}
+ \alpha \sin^{2}\chi_{0}))\\
F_{2}(\theta) &=& \sqrt{\cos^{2}\theta+\alpha \sin^{2}\theta} (1 + \sin^{2}\theta(-\frac{1}{4}
+ \alpha \sin^{2}\chi_{0}))
\eea
We show in Fig. 3 $F_{1}(\theta)$ and $F_{2}(\theta)$ for $\alpha=0.69$ (the value
\begin{figure}[h]
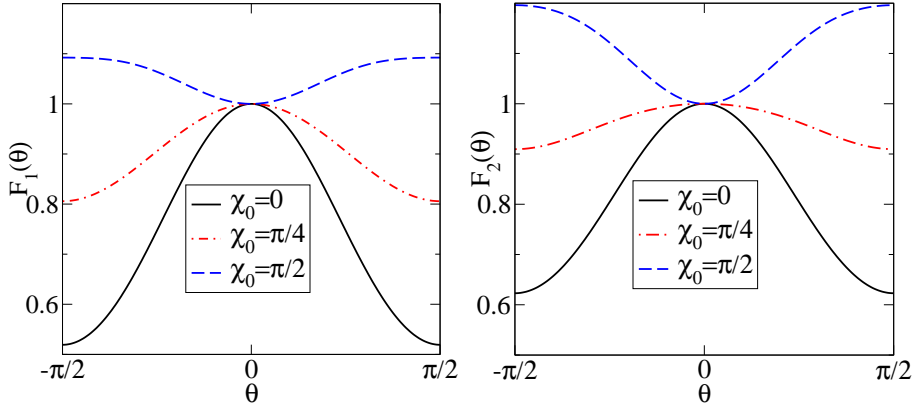

\includegraphics[width=6cm]{fig3a.eps}
\includegraphics[width=5.9cm]{fig3b.eps}
\caption{The angular functions F$_{1}(\theta)$ (left) and F$_{2}(\theta)$}
\end{figure}
appropriate for UPd$_{2}$Al$_{3}$) and $\chi_{0} = 0,\frac{\pi}{4}, $ and $\frac{\pi}{2}$, which is
compared with the experimental data \cite{18} taken at $T=0.4 K$ shown in Fig. 4.  Except
\begin{figure}[h]
\includegraphics[width=6cm]{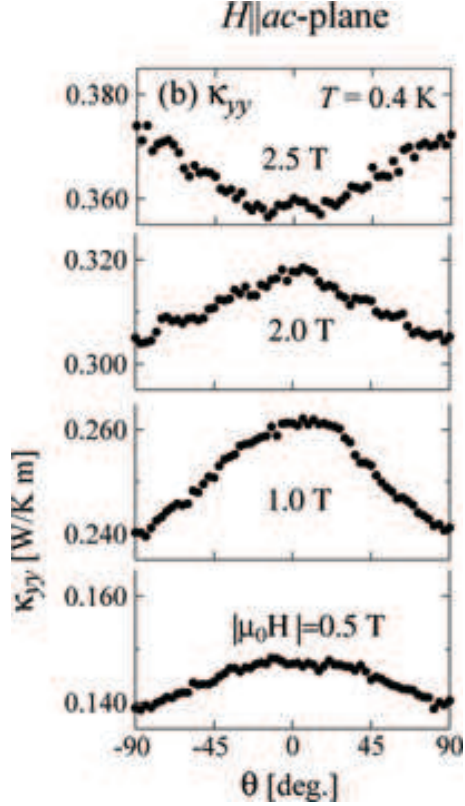}
\caption{Angular-dependent magnetothermal conductivity $\kappa_{yy}$ of UPd$_2$Al$_3$}
\end{figure}
for the data taken for $H=2.5 T$, the data for $H=0.5 T, 1 T$ and $2 T$ are consistent with
$\chi_{0}=\frac{\pi}{4}$, indicating again $f=\cos 2\chi$.  We note also the sign of the twofold
term in $\kappa_{yy}$ at $T= 0.4 K$ changes sign at $H = 0.36 T$.  This is consistent with the
fact that for $T < v\sqrt{eH}$ the nodal excitations are mostly due to the Doppler shift while
for $T > v\sqrt{eH}$ the thermal excitations dominate\cite{24}.

\noindent{\it \bf 4. Concluding Remarks}

We have analyzed recent thermal conductivity data \cite{18} of UPd$_2$Al$_3$ from 
2 perspectives: universal heat conduction and the angle-dependence.  The present study 
indicates $\Delta({\bf k})=\Delta \cos(2\chi)$.  This
is different from the conclusion reached in Ref.\cite{18}. Also we have extended the universal
heat conduction for a class of superconducting order parameters $\Delta({\bf k})$, which
will be useful for identifying the gap symmetry of new superconductors such as URu$_2$Si$_2$ and
UNi$_{2}$Al$_{3}$.

Furthermore, we have worked out the expressions for $\kappa_{yy}$ when the magnetic field is rotated
within the z-x plane.  The angle dependence of $\kappa_{yy}$ is extremely useful to locate
the nodal lines when all nodal lines are horizontal.  Perhaps $\kappa_{yy}$ in Sr$_2$RuO$_4$
will help to identify the precise position of the horizontal nodal lines in $\Delta({\bf k})$, if
a further study of nodal lines is necessary.  Also after UPt$_3$ and UPd$_2$Al$_3$ we expect many
of the U-compound superconducting energy gaps have horizontal lines.

{\bf Acknowledgments}

We thank Balazs Dora, Hae-Young Kee, Peter Thalmeier, and Attila Virosztek for
helpful collaborations on related subjects.

\end{document}